\documentclass[usenatbib,usegraphicx, iop]{emulateapj}
\usepackage{amsmath}
\usepackage[]{hyperref}
\usepackage{graphicx}
\usepackage{color}
\usepackage{units}
\usepackage{ulem}
\usepackage{fixltx2e}
\DeclareGraphicsExtensions{.eps,.pdf,.png,.jpg}
\hypersetup{
    colorlinks=true,
    linkcolor=black,
    citecolor=black,
    filecolor=black,
    urlcolor=black,
}

\shorttitle{Hydrodynamic Shielding}
\shortauthors{Forbes and Lin}

\begin{document}

\title{{\bf Hydrodynamic shielding and the survival of cold streams}}

\author{John C. Forbes}
\affil{Astronomy Department, Harvard University, 60 Garden St., Cambridge MA 02138, USA; \\ john.forbes@cfa.harvard.edu, aloeb@cfa.harvard.edu }

\author{Douglas N. C. Lin}
\affil{Department of Astronomy \& Astrophysics, University of California, Santa Cruz, CA 95064 USA}

\begin{abstract}
Cold clouds in hot media are quickly crushed, shredded, and then accelerated as a result of their interaction with the background gas. The persistence of cold clouds moving at substantial velocities in harsh environments is a common yet puzzling feature of many astrophysical systems, from quasar absorption lines probing galactic halos to clouds of dust passing near Sgr $A^*$. Here we run a set of idealized numerical experiments, subjecting a line of cold clouds at a series of mutual separations to a hot background wind. We find that this stream of clouds is able to shield itself from hydrodynamic destruction by accelerating the hot background material, creating a protective layer of co-moving gas. We write down a simple diffusion equation that reproduces the behavior of the simulations, and we discuss the implications for cosmological gas accretion and G2.
\end{abstract}

\section{Introduction}

In diffuse astrophysical media, it is not uncommon to find cold clouds immersed in a background gas orders of magnitude hotter. This is the natural result of thermal instability in the ISM \citep{field_thermal_1965}, cold accretion in the halos of galaxies \citep{keres_how_2005, dekel_cold_2009}, cooling of galactic winds driven by star formation \citep{wang_cooling_1995,thompson_origin_2016} or AGN \citep{li_modeling_2014}, and a variety of sources proposed to explain the putative gas cloud G2 in the galactic center \citep{gillessen_gas_2012}.

The ubiquity of multi-phase media has encouraged a long history of theoretical studies of idealized cold clouds in hot media. In the simplest case, a single cloud of uniform density is placed in a background medium a factor of $\chi$ hotter and less dense. The cloud and the background are initialized with a relative velocity or such that a shockwave will sweep through the background medium and hence the clouds. This problem is described by two dimensionless numbers, namely the density contrast $\chi$ and the mach number $\mathcal{M}$ of the relative velocity with respect to the sound speed in the background medium \citep{klein_hydrodynamic_1994, pittard_mass-loaded_2006}. Other physics may be relevant depending on the specific problem, including magnetic fields, self-gravity, thermal conduction, internal turbulence, and cooling \citep[e.g.][]{fragile_magnetohydrodynamic_2005, armillotta_efficiency_2016, armillotta_survival_2017, mccourt_magnetized_2015, mccourt_characteristic_2018, banda-barragan_filament_2016, banda-barragan_filament_2018, schneider_cholla_2015, schneider_hydrodynamical_2017}.

The material comprising the cold cloud generically undergoes two separate but related processes as a result of its interaction with the background \citep{klein_hydrodynamic_1994}. First, the cloud is compressed along the direction of the flow by a strong shock, so long as the Mach number of the flow with respect to the cloud's sound speed, namely $\mathcal{M} \sqrt{\chi}$, is large compared to 1. In other words, the cloud is crushed on a timescale of order $t_\mathrm{crush} = r \sqrt{\chi}/v_\mathrm{rel}$, which makes it susceptible to disruption via hydrodynamic instabilities. Second, the cloud experiences drag, slowing its velocity relative to the background gas. The cloud therefore decelerates relative to the background on a timescale of order $t_\mathrm{airmass} = r \chi / v_\mathrm{rel} = \sqrt{\chi} t_\mathrm{crush}$, i.e. the time it takes the cloud to sweep up a mass of background material equal to its own mass.

The ease with which cold clouds can be slowed and disrupted poses problems observationally. Quasar absorption surveys have detected immense quantities of gas with low ionization states in the vicinity of galaxies of all masses and types \citep{tumlinson_large_2011}. If cold clouds can be easily disrupted, the presence of cold gas at large impact parameters (hundreds of kpc) is puzzling. The G2 cloud near the central black hole of the Milky Way survived its recent passage near pericenter, despite the predictions of hydrodynamic simulations that assumed it was a small gas cloud \citep[e.g.][]{pfuhl_galactic_2015}. 

Some additional physics seems to be required to explain these phenomena, and it may be different in each case. In the CGM the cold gas could be continuously condensing out of the hot medium, last for a short time, and then be disrupted. This sort of process can be modeled with a coagulation/disruption equation \citep[e.g.][in the context of star forming cores]{huang_coagulation_2013}. 

Another solution would be to invoke some additional physics to extend the lifetime of the clouds. \citet{murray_dynamical_1993} showed that when a cloud approaches the Bonner-Ebert mass, the disruptive effects it faces are almost completely mitigated by its self-gravity. However, for many of the clouds in question, the temperature of their environments is high enough that the Bonner-Ebert mass is implausibly large. \citet{mccourt_magnetized_2015} have proposed that magnetic fields threading the clouds could suppress the hydrodynamic instabilities and substantially increase their lifetime, while at the same time appreciably increasing the drag they experience.

In this paper we explore the ability a set of clouds traveling together in a stream has to shield itself from hydrodynamic drag and disruption by interaction with the background medium. Generally speaking, this effect should slow the disruption of even purely hydrodynamic clouds, and decrease the drag effect (in contrast to a single magnetized cloud where the drag is increased). In Section \ref{sec:simulations} we describe the setup and results of a simple set of hydrodynamic simulations to demonstrate the effect and its magnitude. In Section \ref{sec:analytic} we use a simple analytic model to gain insight into the simulations results, and finally we discuss the implications for a wide range of multi-phase medium problems at a variety of scales in Section \ref{sec:discussion}.

\section{Simulations}
\label{sec:simulations}
 We employ the publicly-available ENZO adaptive mesh refinement hydrodynamics code \citep{bryan_enzo_2014}. Although ENZO has the capability to directly calculate the effects of magnetohydrodynamics, non-equilibrium radiative cooling, ionizing radiation, gravity, and thermal conduction, we disable all of these features and focus purely on the hydrodynamics. This means that our simulations are essentially scale-free, and in principle the results may be applied anywhere that the density contrast and mach number are comparable to those we use here. On the other hand, there are many scenarios some subset of the additional physics that we have neglected may strongly influence the dynamics.

The code solves the inviscid equations of hydrodynamics. We use a piecewise parabolic mesh interpolation with an HLLC Riemann solver. We also employ a dual-energy formalism: normally the code evolves the total energy and when the internal energy or temperature is needed, it is calculated by subtracting the kinetic energy from the total energy. However, when the gas is highly supersonic (as may be the case in the cloud interiors), the total energy and kinetic energy will be very similar, so the internal energy will be subject to roundoff error. In this case, ENZO switches to evolving an internal energy equation in addition to the total energy equation.

In these simulations, we place a series of clouds a fixed distance apart in a periodic box (in all three dimensions), with each cloud arranged perfectly downstream of the last cloud. We therefore introduce an additional dimensionless number $\delta$, the separation between cloud centers in units of cloud radii. In the following section, we focus on the effect of varying this parameter over a large range, from 2 to 64, while keeping $\chi=100$, and $\mathcal{M}=0.31$ or $1$. These parameters are in the right range for both cold neutral medium clouds interacting with the warm neutral medium, as expected in an interstellar medium subject to classical thermal instability \citep{field_thermal_1965}, and $\sim 10^4 \mathrm{K}$ gas clouds in the hot coronae of Milky Way-mass haloes with Virial temperatures of order $10^6\mathrm{K}$.

The periodic box is initialized with a 16x16x64 root grid with cubic cells, i.e. the simulation box has a 4:1 aspect ratio. The long axis, which we'll call the z-axis, is both the direction of relative motion between the clouds and the background medium and the axis along which the clouds are aligned. The clouds are placed along the z-axis at the center of the box in each x-y plane. Each cloud is spherical with uniform density. The clouds' initial radii are $6.25\%$ of the box's x- or y- size\footnote{ As a grid code, ENZO is not automatically Galilean invariant, so we note that we initialize the clouds to be moving and the background to be stationary in the frame of the computational grid.} For most of the simulations used here, the cloud radius is resolved by 16 cells at the finest AMR resolution, i.e. 4 levels of refinement beyond the root grid are used. Cells are marked for refinement according to the local density gradient - in particular if the value of the density on either side of a cell changes by more than $50\%$ of the cell's density, that cell is marked for refinement. The simulations use an ideal gas equation of state with $\gamma=5/3$, and are run for 200 cloud crossing times, defined as $t_\mathrm{cross} = r/v_\mathrm{rel}$, which for the density contrast adopted here is 20 cloud crushing times, and 2 cloud acceleration times.

\begin{figure*}
\centering
\includegraphics[clip, trim={.1cm 2cm 0cm 1.8cm}, width=13.5cm]{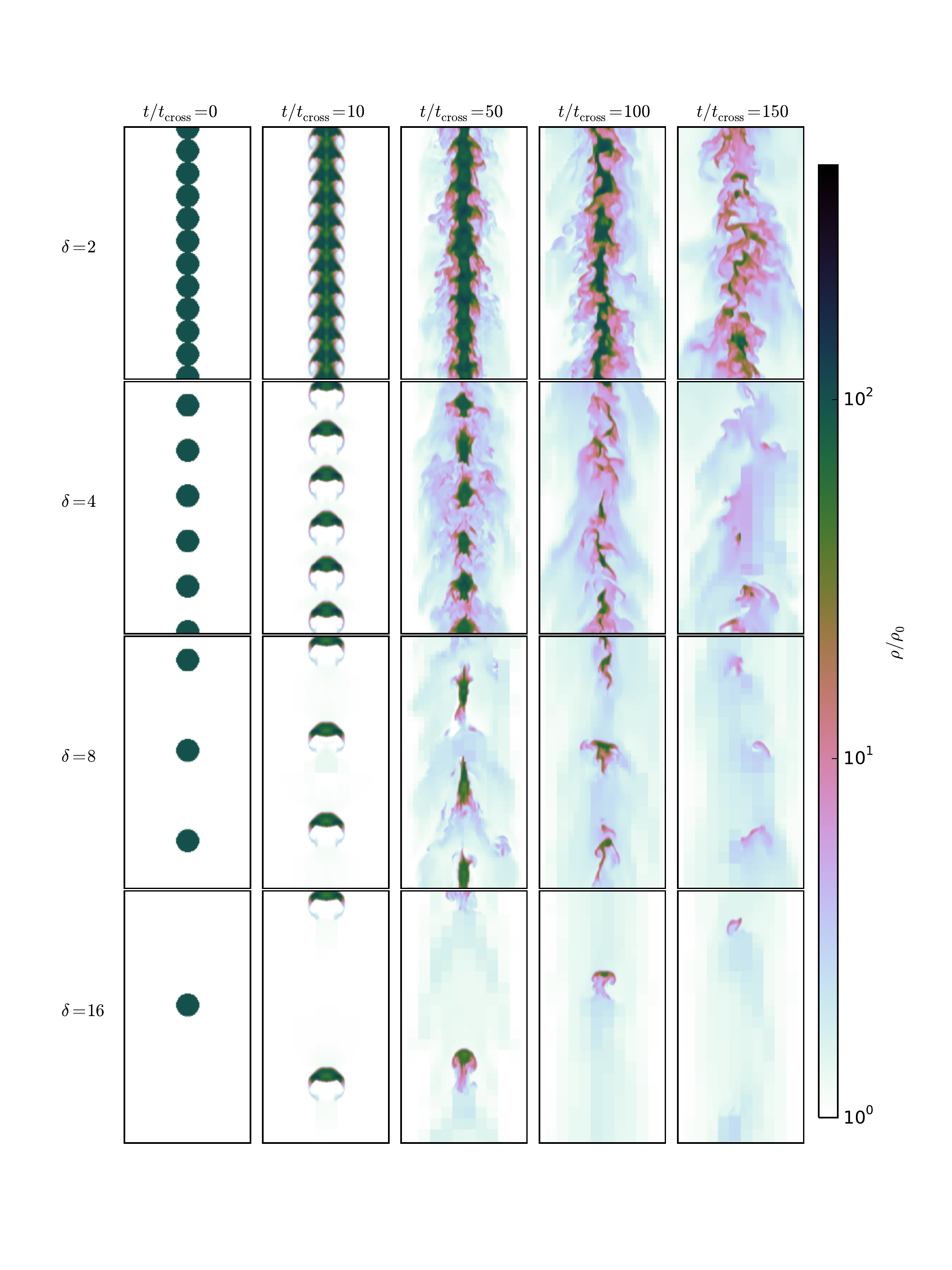}
\caption{Examples of the density structure of four different simulations (rows) at five different times throughout the simulations (columns). These plots are $x$-$z$ slices through the centers of the clouds, with a fixed logarithmic scale in the density. The clouds are initialized with a positive $z-$velocity, i.e. the $y-$axis in these plots.}
\label{fig:density}
\end{figure*}

Figure \ref{fig:density} shows the gas density relative to the initial background density $\rho_0$ in slices  through a subset of the simulation domain for the $\mathcal{M}=1$ simulations at a few different values of the separation of the cloud centers $\delta$. The first two columns show the evolution of the clouds after a single cloud crushing time, and indeed, the clouds are flattened and deformed almost regardless of the initial cloud separation. After five crushing times (the third column), the symmetry of the clouds is lost, and substantial differences between the cases with different values of $\delta$ appear. The clouds have also lost a substantial amount of mass owing to hydrodynamic shredding; even though the cloud cores remain intact, the density of material a few cloud radii away both between the clouds and perpendicular to the axis of motions is now a few times higher than the initial background density. After one acceleration timescale, gas comparable to the initial density $\rho=100\rho_0$ has all but disappeared from the more ``isolated'' clouds, with $\delta \ga 16$, but persists in the more closely-spaced clouds. By 1.5 acceleration times, the clouds have lost so much mass that the maximum density contrast in the simulation is closer to 10 than the initial value of 100, except for the closest-spaced case $\delta=2$.

\begin{figure}
\includegraphics[width=9.5cm]{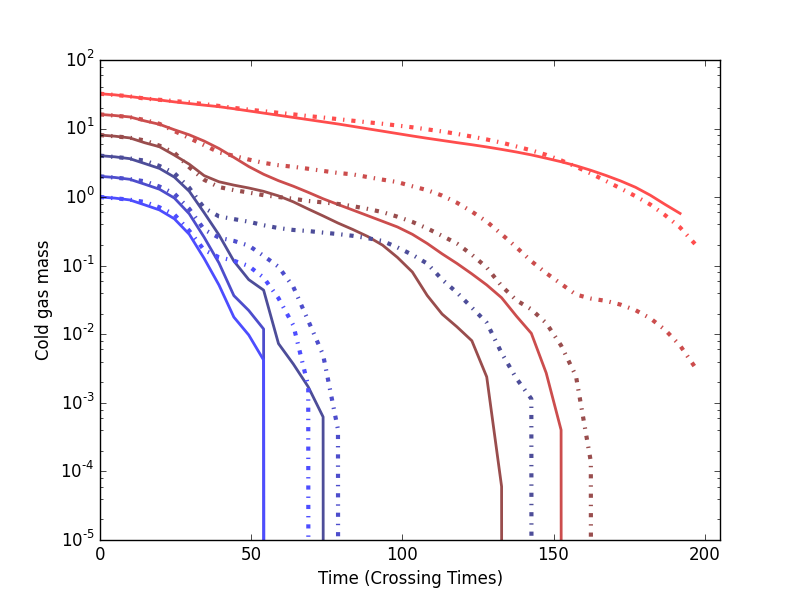}
\caption{The mass of cold material in each simulation in units of the mass of a single cloud. Lines from red to blue indicate increasing cloud separations by powers of two: $\delta = 2,4,8,16,32,64$, and solid vs dot-dashed lines indicated $\mathcal{M}=1$ and $\mathcal{M}=0.31$ respectively. }
\label{fig:coldmass}
\end{figure}

\begin{figure}
\includegraphics[width=9.5cm]{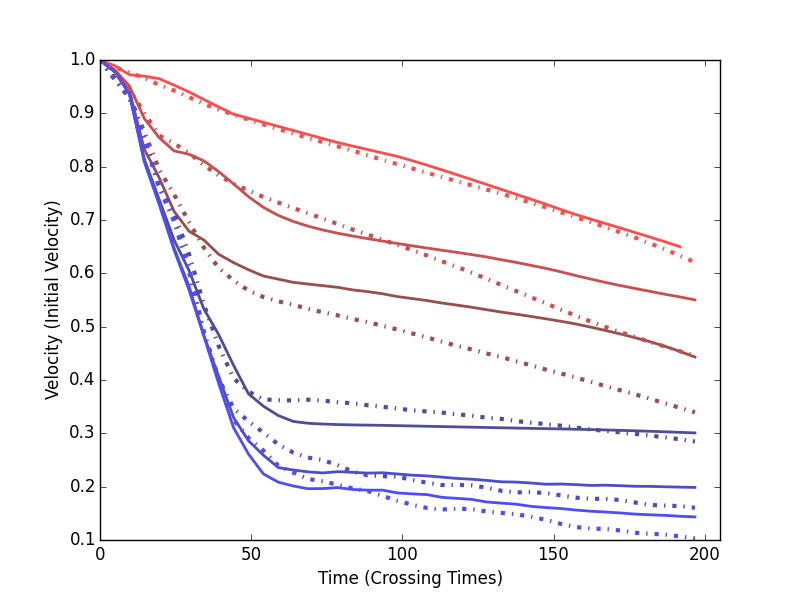}
\caption{The velocity of the cloud material over time. In these units and in this frame, the clouds initially have a velocity of 1 and the background gas has a velocity of zero. As in Figure \ref{fig:coldmass}, lines from red to blue indicate increasing cloud separations by powers of two: $\delta = 2,4,8,16,32,64$, and solid vs dot-dashed lines indicated $\mathcal{M}=1$ and $\mathcal{M}=0.31$ respectively..}
\label{fig:velocity}
\end{figure}

These trends are borne out more quantitatively by Figures \ref{fig:coldmass} and \ref{fig:velocity}, showing respectively the mass of cold gas and the velocity of material associated with the initial clouds. The cold gas mass is normalized to the mass of a single cloud, and defined as the mass of material whose temperature is below $10\ T_c$, where $T_c$ is the initial temperature of the cold gas. Because the simulations are initialized in thermal pressure equilibrium with a density contrast of 100, this corresponds to a temperature logarithmically halfway between the initial temperatures of the cold and hot medium. 

As we saw based on the morphology of the clouds in Figure \ref{fig:density}, the mass of cold gas rapidly decreases over time, with an inflection point at a few $t_\mathrm{crush}$ for clouds with $\delta \ga 16$. This marks a transition from a regime in which clouds are hydrodynamically disrupted individually to one where they can protect each other to some degree. Since clouds travel a distance of $r_c\sqrt{\chi}$ in a cloud crushing time, for these simulations where $\chi=100$ it makes sense that this transition occurs around a $\delta$ of 10. Clouds spaced much farther apart will not interact with gas that has interacted directly with the preceding cloud before they are individually shredded. 

The dot-dashed lines in Figures \ref{fig:coldmass} and \ref{fig:velocity} indicate the results for $\mathcal{M}=0.31$. The cold gas mass is similar to the $\mathcal{M}=1$ case (solid lines) until a few times $t_\mathrm{crush}$, at which point the two cases diverge slightly. The slower relative velocities allow more cold gas to survive somewhat longer, and also differ in the transition value of $\delta$ between the clouds being individually disrupted vs. mutually shielding each other. This makes sense because a cloud does not need to reach the position of its preceding cloud to be affected by the preceding cloud; it merely needs to reach the position at which information about the preceding cloud has propagated through the background medium. Setting the time it takes for the signal from the first cloud to reach the second equal to the cloud crushing time implies that 
\begin{equation}
\delta_\mathrm{shield} = (1+\mathcal{M}^{-1}) \sqrt{\chi}
\label{eq:deltashield}
\end{equation}
approximately marks the transition between the two regimes. For our two sets of experiments $\mathcal{M}=1$ and $\mathcal{M}=0.31$, the right hand side evaluates to $20$ and $40$ respectively, in reasonable agreement with the simulations where the boundary appears at $\sim 16$ and $\sim 32$ respectively. 

Figure \ref{fig:velocity} shows the mass-weighted mean velocities in the $z-$direction of the gas originally associated with the cold medium, followed over the course of the simulation by a passive tracer field, in the initial rest frame of the background gas. These may be interpreted as the average velocities of the cold components for as long as a substantial fraction of the initial gas mass remains cold. Once the clouds in a given simulation have been disrupted by hydrodynamic instabilities, this velocity traces the residual lower-density stream of material. 

The most obvious trend in Figure \ref{fig:velocity} is that the rate at which the clouds are slowed by their interaction with the background gas monotonically increases with $\delta$. That is, the closer together the clouds are, the longer it takes them to slow down. The clouds that are completely disrupted after a few cloud crushing times all behave similarly, with a steep initial drop in velocity, falling to about $1/3$ of their initial velocity in the time it takes them to be shredded. For clouds that remain intact, the time it takes to slow them down by a similar amount exceeds $t_\mathrm{accel}$. 

Recall that the expectation for an isolated cloud is that its velocity should change on a timescale of order $t_\mathrm{accel} = t_\mathrm{crush}\sqrt{\chi} = t_\mathrm{cross}\chi$. None of the simulations quite fit this expectation, but in retrospect this should not be surprising because none of the simulations have an isolated unchanging cloud being accelerated. Clouds in the large-$\delta$ simulations are being disrupted, so the material initially associated with those clouds is being mixed into the background flow, decelerating it somewhat faster than if the clouds had remained intact. Meanwhile the low-$\delta$ simulations, which we have already shown allow the clouds to avoid rapid hydrodynamical disruption, also mutually reduce the drag on each other, thereby extending the time it takes to decelerate the cloud material. We also note that, even in the cases where the clouds are totally disrupted in a few crushing times, the asymptotic velocity of the cloud material never reaches the initial velocity of the background gas (zero in this frame) - this will be discussed further in the following section.

\section{Physical Interpretations}
\label{sec:analytic}

In the previous section, we saw that closely-spaced clouds can behave strikingly differently from the standard isolated cloud case. Closely spaced clouds have smaller accelerations relative to the background medium, resist disruption for longer, and they have a non-zero asymptotic velocity relative to the background medium. In our simulations, we systematically varied the initial separation of the clouds, as well as the resolution and the initial velocity. In this section, we attempt to gain some insight into these results.

\subsection{The cylinder model}

As a set of cold clouds are subject to hydrodynamic disruption by interaction with the hot background medium, we can imagine two distinct outcomes after a large amount of time has passed. The clouds can cool efficiently and maintain something similar to a multi-phase medium, where roughly speaking a population of clouds would exist at a similar temperature and density to the initial clouds, but perhaps with smaller radii \citep[e.g. set by the the cloud shattering radius of][]{mccourt_characteristic_2018}, or some distribution of radii \citep[e.g.][]{lin_two-phase_2000, huang_coagulation_2013}. Alternatively, if the clouds cannot cool or be held together by another force, they will be completely disrupted and mixed with the background medium. In both cases, we can imagine that the mixture region is a cylinder of radius $\xi r_c$. If the cloud centers are separated by a distance $\delta r_c$, we have enough information to compute the long-run state of the system assuming that the cylinder itself is subject to minimal drag and hydrodynamic disruption.

Of these two possibilities, the multi-phase medium outcome is more straightforward to compute. In the frame of the background medium, the initial momentum of the system over a given distance along the line of clouds is $p_0 = N_c m_c v_0$, where $N_c$ is the number of clouds in the volume, $m_c$ is the mass of one cloud, and $v_0$ is the initial relative velocity. In the final state, the cold material has momentum $N_c m_c v_f$, where $v_f$ is the final velocity. Part of the momentum has been transferred to the cylinder of hot material, which we envision as a cylindrical pillbox around each cloud, so the momentum from this component is $ \rho_h N_c v_f (\pi (\xi r_c)^2  \delta r_c - (4/3)\pi r_c^3)$. Summing the two final-state momenta and equating them to the initial momentum, we can solve for the final velocity as a function of the initial velocity
\begin{equation}
\label{eq:velocity}
\frac{v_f}{v_i}  = \frac{\chi}{\chi + (3/4)\xi^2\delta -1}
\end{equation}

Meanwhile in the case where cooling is inefficient, the cylinder of mixed material will generically have a different density than the initial value, so in addition to $v_f$, we will have to solve for $\rho_\mathrm{mix}$, the density of the mixture. To do so we will employ mass conservation as well as momentum. The initial mass of the system is $N_c m_c + \pi\rho_h r_c^3 N_c (\xi^2 \delta - 4/3)$, while the final mass is $\pi \rho_\mathrm{mix} r_c^3 \xi^2\delta N_c$. The ratio of the mixture density to the background density is therefore
\begin{equation}
\label{eq:rhomix}
\frac{\rho_\mathrm{mix}}{\rho_h} = 1 + \frac{4}{3\xi^2\delta}(\chi-1)
\end{equation}
With this result, we can compute the ratio of the final velocity to the initial velocity by setting the initial momentum $N_c m_c v_0$ equal to $\pi \rho_\mathrm{mix} \delta \xi^2 r_c^3 N_c v_f$. The result is in fact identical to the efficient-cooling case, Equation \ref{eq:velocity}.

\begin{figure}
\includegraphics[width=9.5cm]{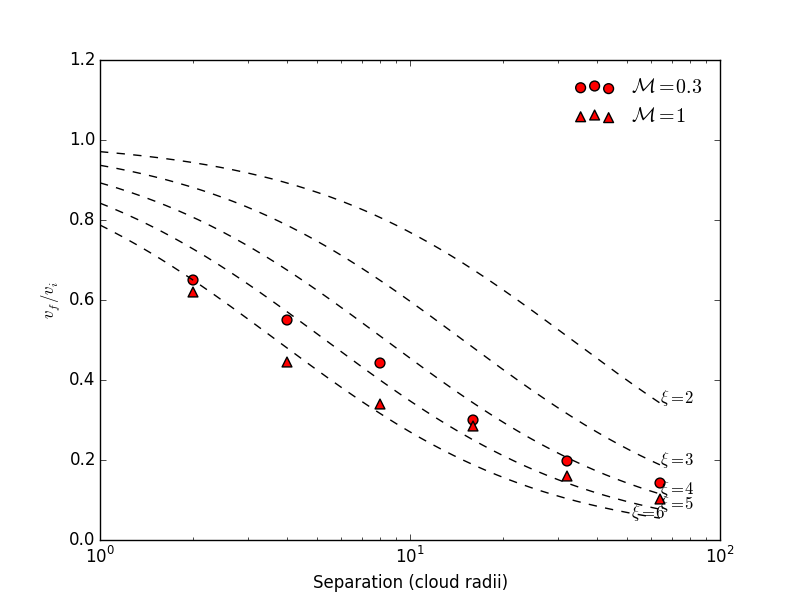}
\caption{The final velocity of cloud material in the simulations at $t=200 t_\mathrm{cross}$. For comparison, lines from the cylinder model discussed below are presented for different values of $\xi$.}
\label{fig:xi}
\end{figure}

This leads us to the conclusion that regardless of the state of the gas, we have a reasonable estimate of its velocity provided we can estimate $\xi$, $\delta$, and $\chi$. For our simulations, $\delta$ and $\chi$ are inputs, so we can measure $v_f/v_0$ at the end of each simulation and plot the result as a function of $\delta$, which we do in Figure \ref{fig:xi}. The result suggests that $\xi$ lies between 4 and 6. There are a number of caveats, however. First, we can see from Figure \ref{fig:velocity} that the simulations with the closest spacings still have steadily decreasing velocities at $t=200 t_\mathrm{cross}$, while the more isolated clouds have reached something closer to a steady state. This suggests that the leftmost points in Figure \ref{fig:xi} are overestimated. Moreover, values of $\xi$ approaching 8 are cause for concern, because that is the size of the computational box. Essentially one should worry that the stream is being influenced by the boundary, and it may not be treatable as isolated.

\begin{figure*}
\includegraphics[width=17cm]{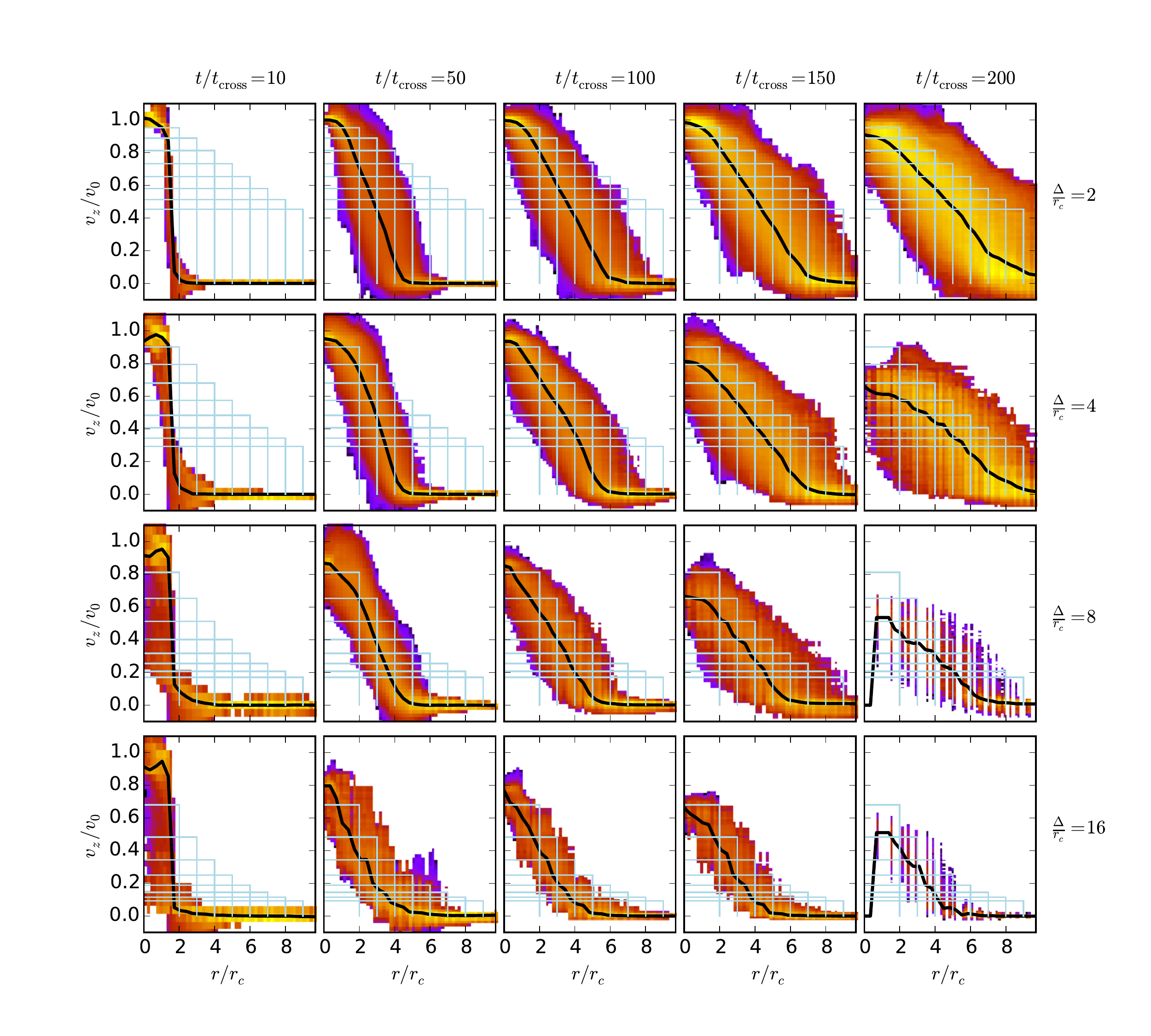}
\caption{The mass-weighted velocity as a function of radius for several separations (rows) and times(columns). Light blue lines show predictions for the simple cylinder model discussed in the text. The black line shows the mass-weighted median velocity at each radius. The background histogram shows the mass of material in that pixel, scaled logarithmically from white/purple to yellow.}
\label{fig:r_vz}
\end{figure*}

\begin{figure*}
\includegraphics[width=17cm]{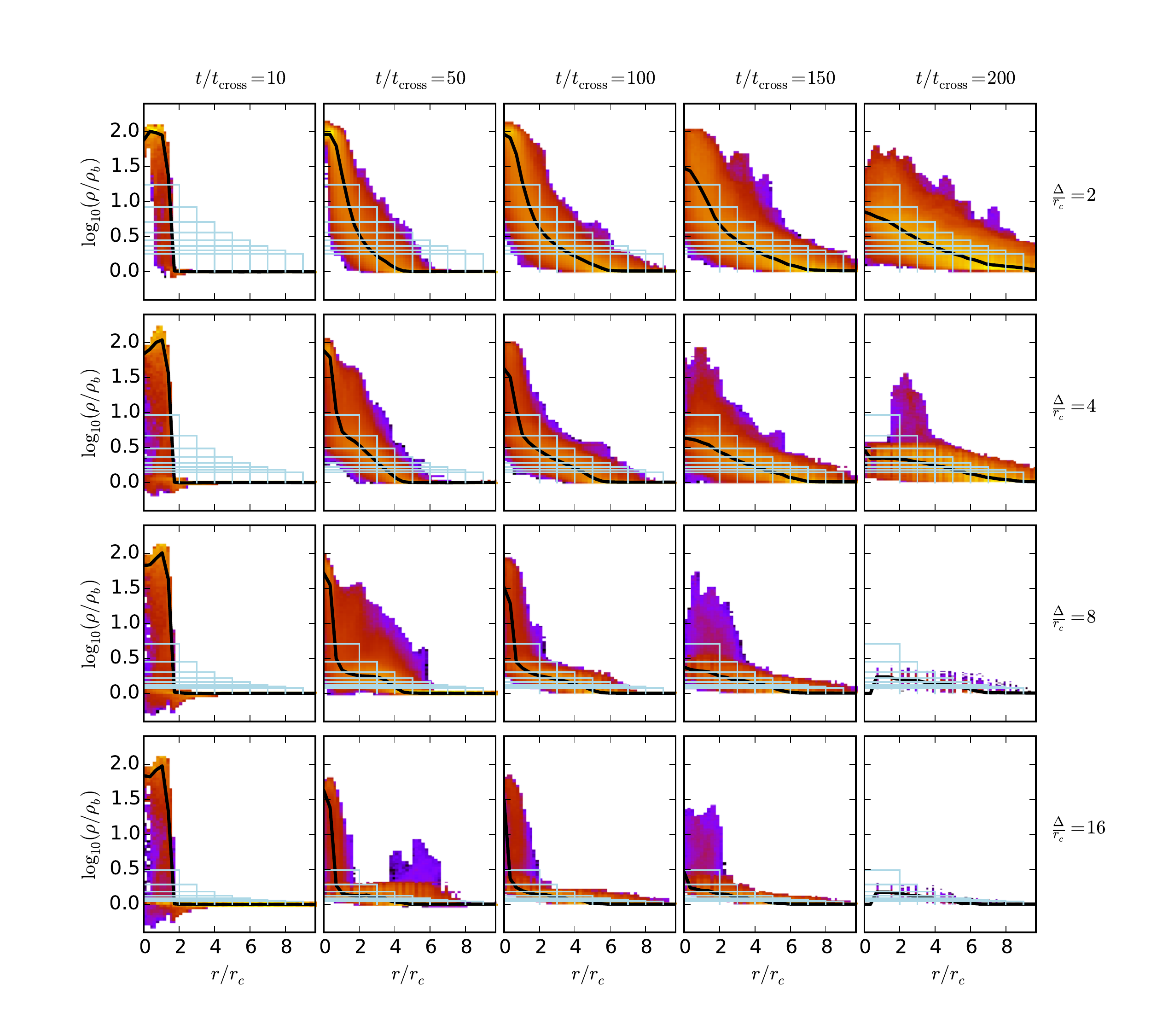}
\caption{The mass-weighted density as a function of radius for several separations (rows) and times(columns). Light blue lines show predictions for the simple cylinder model discussed in the text. The black line shows the mass-weighted median density at each radius.}
\label{fig:r_rho}
\end{figure*}

We can explicitly check how true this simple approximation is in the simulations themselves. Figures \ref{fig:r_vz} and \ref{fig:r_rho} show mass-weighted maps of $r$ vs $v_z$ and $r$ vs $\rho$ for a variety of separations (rows) and times (columns). The light blue rectangles show our expectations for $v_z$ and $\rho$ respectively as a function of radius - essentially we would expect the gas to fall close to one of these lines, and in particular, given Figure \ref{fig:xi}, we would expect the gas to follow one of the rectangles ending at $r/r_c$ between 4 and 6. However, it is clear that the simulation follows a much more tapered profile, i.e. a transition from an interior region with a large velocity to an exterior region with near-zero velocity, that gets wider with time. 

\subsection{Diffusion Equation}

The median velocity profiles shown as black lines in figure \ref{fig:r_vz} give a hint of how to proceed. Although the simulations all begin with cold clouds in a hot medium at different velocities, very quickly the velocity distribution shown in the colormaps coalesces around the median profile. Even though the gas is still in a sense multiphase, exhibiting order-of-magnitude variation in density even at fairly late times (see Figure \ref{fig:r_rho}), it has a well-defined, largely single-valued velocity profile. This leads us to the conclusion that the kinematics of the system are largely governed not so much by the interaction of the cold clouds with the hot background, e.g. clouds traveling in each other's wakes, but rather by a diffusion process wherein the initial momentum in the center of the profile is smoothly diffused outwards by some viscosity.

We can model this process by writing down the equations of hydrodynamics in cylindrical coordinates (with the $z-$axis aligned with the direction of the flow) with the approximation that all derivatives with respect to $\phi$ or $z$ are zero. In this limit, the evolution of the $z-$velocity is described by
\begin{equation}
\frac{\partial u_z}{\partial t} = -u_r \frac{\partial u_z}{\partial r} + \frac{1}{r\rho} \frac{\partial}{\partial r}\left( r\mu \frac{\partial u_z}{\partial r}\right),
\end{equation}
where $u$ is the fluid velocity, $\rho$ is the density, and $\mu$ is the dynamic viscosity. We proceed to solve this equation on a grid of regularly-spaced $r-$values from $r=0$ to $r=R=8$ in units of initial cloud radii, subject to the boundary conditions $\partial u_z/\partial_r |_{r=0} = \partial u_z/\partial_r |_{r=R}= 0$ and initial conditions drawn from the mass-weighted median values of these quantities at a given radius in the simulation.

\begin{figure}
\includegraphics[width=9cm]{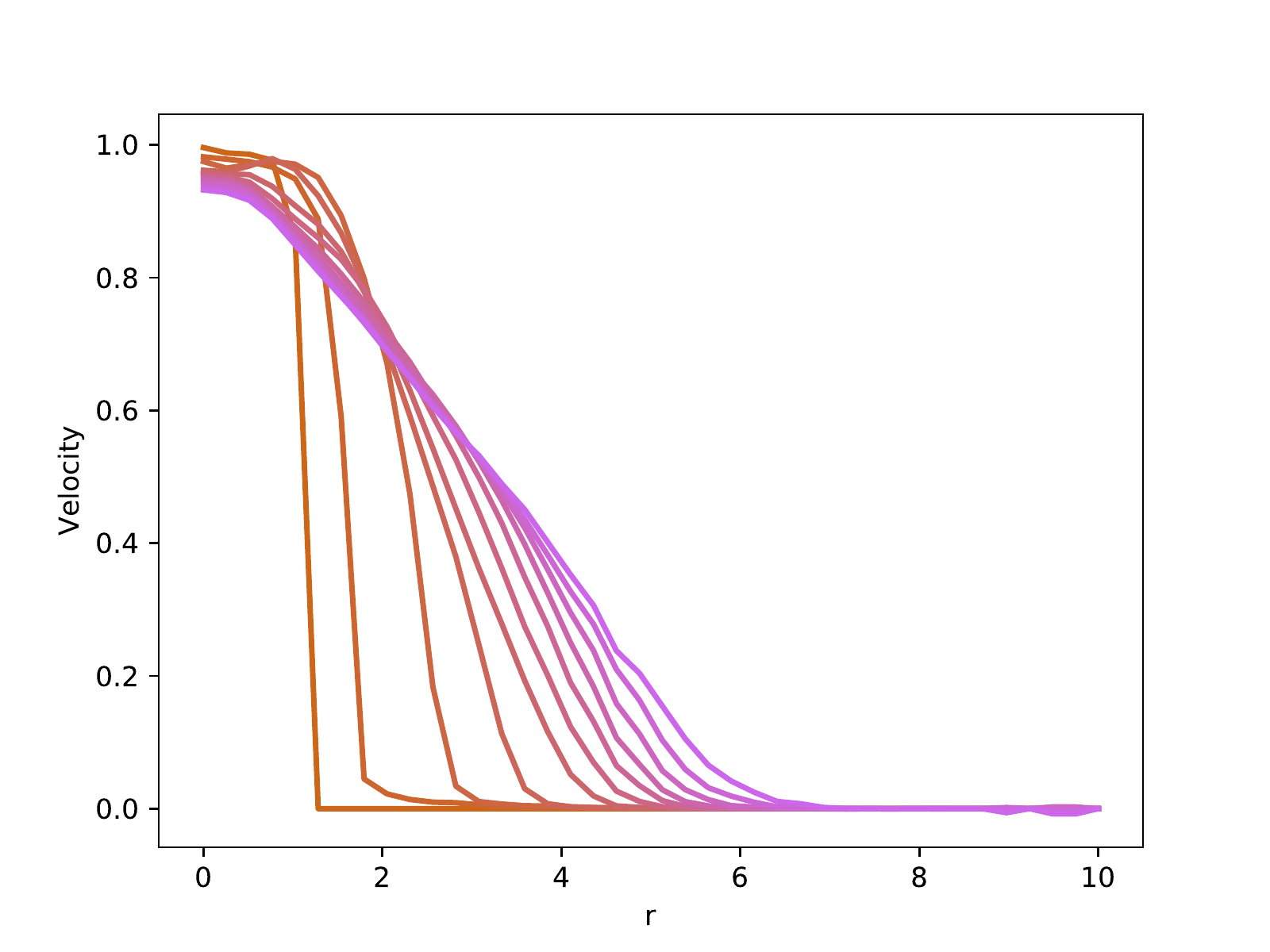}
\caption{The median velocity profile from the 3D simulation. The median velocity at each radius (analogous to the black lines in Figure \ref{fig:r_vz}) in the $z-$direction extracted from the $\mathcal{M}=1$, $\delta=4 $ simulation is shown in units of the initial velocity $v_0$ as a function of radius in units of $r_c$. Each line shows the velocity at a different time, with red lines indicating earlier, and purple lines indicating later times. A line is plotted for every 10 crossing times.}
\label{fig:uzsim}
\end{figure}

\begin{figure}
\includegraphics[width=9cm]{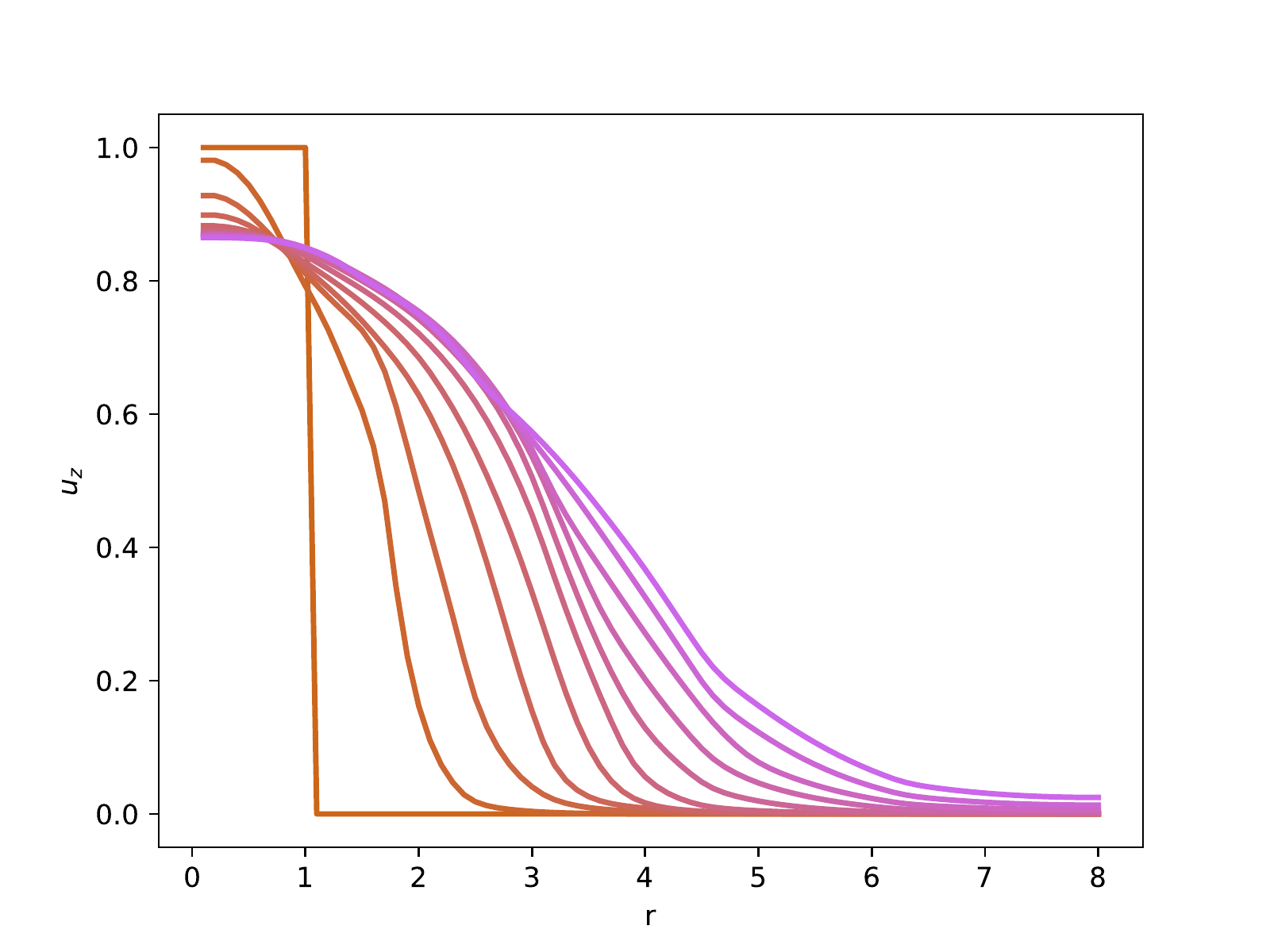}
\caption{The velocity profile from the viscous model tuned to fit the results of Figure \ref{fig:uzsim}. }
\label{fig:uzmodel}
\end{figure}

\begin{figure}
\includegraphics[width=9.5cm]{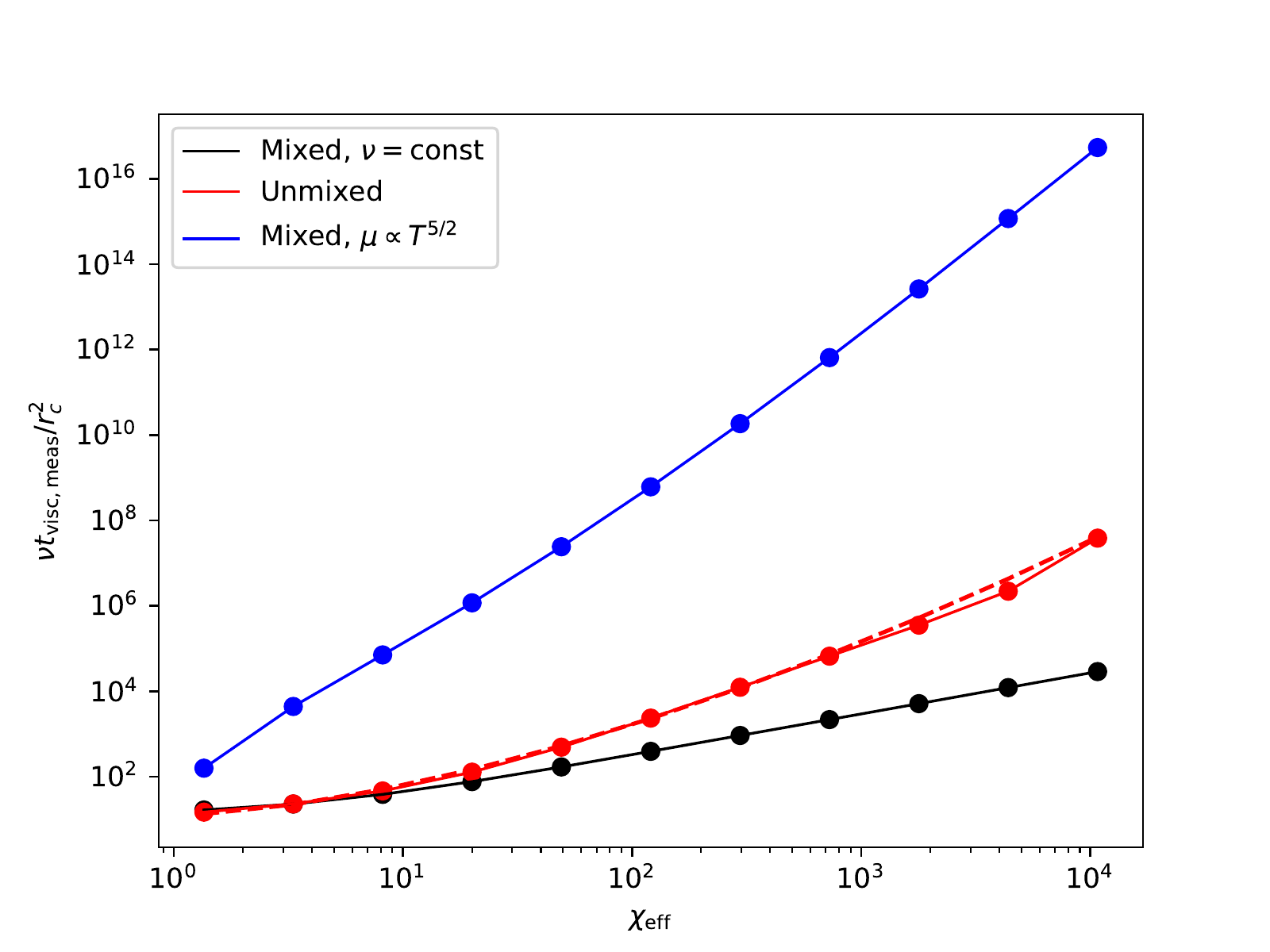}
\caption{The viscous timescale. Here we plot the viscous timescale relative to the simple estimate thereof $r_c^2/\nu$ as a function of the effective over-density of the stream of clouds. }
\label{fig:tvisc}
\end{figure}

Equations for $\rho$ and $u_r$ are straightforward to derive, whereas $\mu$ is typically taken to be a fundamental property of the fluid. At an order of magnitude level, $\mu \sim v \lambda/\rho$, where $v$ is the typical velocity of the intermediary diffusing momentum (e.g. the sound speed for molecular viscosity), and $\lambda$ is the typical length scale of the diffusion process (e.g. the mean free path of the particles). In the numerical simulations, $\lambda$ should be of order the local cell size. As a first step, we consider the case where $\rho$, $u_r$, and $\lambda$ are given by their mean values at a given radius and time in the 3D simulations (values are sampled every 10 crossing times at 40 evenly-spaced radii between 0 and 10 cloud radii). We then assume that $\mu = \mu_0 \lambda v_0 \rho$, and adjust $\mu_0$ until the profiles of $u_z$ on our 1D mesh reasonably reproduce the the mean values of $v_z$ measured in the 3D simulations. We find that substantially better agreement may be obtained if we are also free to adjust $u_r$ as extracted from the simulations by a constant factor $f_{u_r}$; using the measured mean $u_r$ rapidly advects much of the momentum from the stream outwards, yielding $u_z$ profiles much broader than we measure. The best agreement is obtained for $\mu_0 = 0.11$ and $f_{u_r} = 0.26$. Figures \ref{fig:uzsim} and $\ref{fig:uzmodel}$ show the velocity profiles for our $\mathcal{M}=1$, $\Delta = 4 r_c $ simulation and the corresponding viscous evolution model respectively.

\section{Discussion}
\label{sec:discussion}

Based on the previous section, it is clear that the primary phenomena we are seeing in our simulations is that of viscous evolution. We argue that individual clouds may be disrupted or not, precipitated, or re-formed depending on the microphysics relevant to the particular problem, but the dynamical process we have documented here should be generic. In order to incorporate it into the classical picture of cloud disruption and acceleration discussed in the introduction, we propose that in addition to $t_\mathrm{cross}$, $t_\mathrm{crush}$, and $t_\mathrm{airmass}$, one should consider a viscous timescale for the series of clouds, which may be quite different from $t_\mathrm{airmass}$, and more relevant in cases where multiple clouds move coherently together. In what follows, we quantify the viscous timescale for a few possible behaviors of the viscosity, place limits on the cases where the viscous time is more relevant than the classic $t_\mathrm{airmass}$, and apply these results to a few astrophysical examples. 

\subsection{The viscous timescale}
A simple estimate of the viscous timescale based on dimensional analysis is $t_\mathrm{visc} \sim r_c^2/\nu$. We can employ the simple 1D scheme described in the previous section to estimate the viscous timescale for a variety of circumstances by measuring the timescale over which the velocity at $r=0$ approaches the velocity of the background. Explicitly, the viscous timescale we measure is 
\begin{equation}
\label{eq:tvisc}
\frac{t_\mathrm{visc, meas}}{t_\mathrm{cross}} = \frac{100}{\ln(v_{100}/v_{200})}
\end{equation}
Here $v_{100}$ and $v_{200}$ are the velocities of the centermost cell in the evolution of the 1D model, in the frame where the background gas is initially at rest, at $t/t_\mathrm{cross}=100$ and $200$ respectively. This is simply the e-folding time in the case that the velocity is decreasing exponentially. 

When we were employing the 1D code to reproduce the results of the 3D simulations, we estimated the characteristic length scale of the viscous process to be proportional to the average cell size in the particular simulations we were reproducing. Now that we are using this model to make independent predictions, we need not be tied to the limitations of the 3D simulations. Instead, we can specify some estimate for how the viscosity should vary in a real-world application. Here we can envision a few possibilities. 

It could be the case that in some systems, just like classical accretion disks, the dominant viscosity may ``anomalous.'' In such cases, following \citet{shakura_black_1973} one could envision setting $\nu = \mu/\rho = \alpha \lambda v$, where $v$ is the characteristic velocity, and $\lambda$ is a characteristic distance. Reasonable guesses for the problem we are considering here would be $\lambda\approx r_c$, particularly if the same mechanism causing the effective viscosity is responsible for setting the size of the clouds themselves, and $v\approx v_0$. The parameter $\alpha$ is free, but should remain $\la 1$.

The ordinary viscosity in high-temperature plasmas can be substantial owing to its steep $T^{5/2}$ temperature dependence \citep{spitzer_physics_1962, sarazin_x-ray_1988}. Following \citet[e.g.][]{roediger_kelvin-helmholtz_2013}, we include a prefactor $f_\mu<1$ to account for possible suppression by the geometry of the magnetic field, so that
\begin{equation}
\label{eq:spitzer}
\mu = 5500\ \mathrm{g}\ \mathrm{cm}^{-1}\mathrm{s}^{-1} f_\mu (T/10^8 K)^{5/2}
\end{equation} 
This viscosity may be incorporated into our 1D model in one of several ways. First, if the clouds in the stream remain intact, via the action of magnetic fields, rapid cooling, hydrodynamic shielding, or yet something else, then the viscous force is really being mediated through the background gas between the clouds and the background gas far from the stream. In this case, the relevant viscosity is constant as a function of radius. If, however, the individual clouds are disrupted and mixed into a single-phase stream, then we might expect that the thermal pressure is roughly constant as a function of radius, so that $\mu \propto \rho^{-5/2}$.

For each of these cases:
\begin{enumerate}
\item Surviving clouds (unmixed), for which $\mu$ is constant as a function of radius.
\item Disrupted well-mixed clouds, where the value of $\nu$ is assumed to be constant, corresponding to anomalous viscosity
\item Disrupted well-mixed clouds, where the value of $\mu$ is assume to obey the Spitzer formula, i.e. $\mu \propto T^{5/2} \propto \rho^{-5/2}$
\end{enumerate}
we run a series of models for different values of the effective density contrast, namely the $z-$averaged over-density of the stream of clouds relative to the background. Explicitly this effective overdensity $\chi_\mathrm{eff}$ is identical to the density contrast of the mixture defined by equation \ref{eq:rhomix}, where $\xi \approx 1$ if the clouds are unperturbed, i.e.
\begin{equation}
\chi_\mathrm{eff} = 1 + \frac{4}{3^2\delta}(\chi-1).
\end{equation}
For each case we evaluate the viscous time using equation \eqref{eq:tvisc} and normalize it to $r_c^2/\nu$, our order of magnitude estimate for the viscous time, where $\nu$ is evaluated in the background material; depending on which of the three viscosity laws we use, $\nu$ may be substantially smaller in the stream. The results of these measurements are shown in Figure \ref{fig:tvisc}, along with the line
\begin{equation}
\label{eq:tviscFit}
\log_{10}\left( \frac{t_\mathrm{visc,meas}}{r_c^2/\nu} \right) = \log_{10} f(\chi_\mathrm{eff}) = {1.1+0.7 (\log_{10}\chi_\mathrm{eff})^{1.6} }
\end{equation}
which will serve as a rough but useful fit to the unmixed case.

Clearly viscosity requires somewhat more time to act on the clouds than our initial guess based on the cloud size and background viscosity. This offset is consistent with the fact the clouds may quickly accelerate the background material out to a few cloud radii perpendicular to the direction of relative motion. Also of note is the substantial dependence on the effective density - the larger the density of the stream, the longer it takes to accelerate the clouds. This makes sense qualitatively, since the stream's inertia increases along with this density.

\subsection{Limitations}

So far we have demonstrated that the timescale to accelerate cold clouds is the viscous timescale rather than the airmass timescale in an idealized setup where an infinite series of clouds are spaced evenly one after another along the axis of motion in a stationary spatially constant background. In this section we place some basic conditions on where we expect our results to hold.

First, we require that the clouds are close enough together that they can influence each other before they are decelerated in an airmass time. This is equivalent to requiring that the mass of background material between the clouds cannot exceed the mass of material in the cloud itself, or roughly $\chi \ga \delta$. In analogy to equation \eqref{eq:deltashield}, we can include the fact that information about one cloud can propagate in the background medium at its sound speed, in which case the critical separation at which clouds may still conceivably act dynamically as a stream is of order $\delta_\mathrm{accel} = (1+\mathcal{M}^{-1}\chi)$

The picture of hierarchical crossing, crushing, and acceleration timescales implicitly requires that the shock passing through the cloud is strong, i.e. that $\mathcal{M} \ga 1/\sqrt{\chi}$. We also require that the angle between the clouds and the axis of motion is not too large. Since in terms of the simplified cylindrical model introduced earlier, we find $\xi \sim 5$, we expect that when $\tan \theta \la 5/\delta$, the clouds may still act coherently as a stream.

In realistic environments, the background material is likely to be stratified, and itself turbulent. Pressure gradients in the background gas, e.g. to support itself thermally against gravity in an atmosphere, will likely impede the development of large velocities in the background gas outside of the stream. We therefore expect that, over a pressure scale height $H$, clouds in the stream should not be too far apart, i.e. $\delta r_c \la H$, or at the very least $\chi r_c \la H$.

\subsection{Examples}
Here we run through a few cases where mutual hydrodynamical shielding of cold gas in a hot background may play a substantial role. True simulations of each of these individual cases is beyond the scope of this work, and in many cases involves a great deal of non-trivial microphysics. Nonetheless, the physical process we propose here may be at play as well.

\subsubsection{G2}
In 2012 the discovery of a putative gas cloud, dubbed G2, on an orbit consistent with an imminent close pericenter passage of the Milky Way's supermassive black hole \citep{gillessen_gas_2012} prompted a number of genuine predictions for its behavior during and after pericenter \citep[e.g.][]{anninos_three-dimensional_2012, burkert_physics_2012, murray-clay_disruption_2012}. G2 seems to have survived its passage near the black hole with little change in its Br$\gamma$ emission or substantial flares from the black hole itself \citep{pfuhl_galactic_2015, bower_radio_2015}. While the cloud was not hydrodynamically disrupted, in contrast to some predictions, a \citet{pfuhl_galactic_2015} claim that G2's trajectory shows deviations from a simple Keplerian orbit indicating some interaction with the background gas.

Of particular interest to this work is the existence of another putative gas cloud, G1 \citep{clenet_dual_2005}, on an orbit similar to, but likely distinct from, G2's orbit \citep{phifer_keck_2013, pfuhl_galactic_2015, witzel_post-periapsis_2017}. G1's trajectory places it about a decade ahead of G2, leading some authors to speculate that G2 will follow G1's path, with the difference in the two clouds' Keplerian orbits being explained by gas drag \citep{mccourt_going_2016,madigan_using_2017}. 

Models for the evolution of G2 \citep[e.g.][]{steinberg_probing_2018} are consistent with a density contrast between the cold and hot medium of $\chi \ga 10^3$, both in the initial conditions and the subsequent evolution of the cloud along its trajectory. This is consistent with the expectation that the cloud has a temperature of order $10^4 K$, and the background, following e.g.  \citet{yuan_nonthermal_2003} has a temperature between $10^7$ K and $10^8$ K over the course of G2's modeled trajectory, though very close to pericenter, if $T \propto r^{-1}$, the temperature may approach $10^9$ K.

Now that we have an estimate for the density contrast, we would like to estimate the various timescales in the problem in the language of clouds and streams that we have used throughout this work. There is some ambiguity in applying the same concepts to this situation, since the background medium is likely stratified, and the cloud follows an extreme orbit, as opposed to a straight path through a homogeneous medium. For now, we will focus our attention on pericenter, since this is where the cloud is likely to undergo its greatest hydrodynamic drag. Along a Keplerian orbit, the orbital velocity is $\sqrt{GM(2/r - 2/a)}$. For G2's orbit as quoted by \citet{pfuhl_galactic_2015}, the pericenter velocity is about $7900 \mathrm{km}\ \mathrm{s}^{-1}$.

Taking the cloud's radius to be $\sim 1.7 \times 10^{15}\mathrm{cm}$, we estimate the crossing time, crushing time, and airmass time of the cloud to be, respectively, $t_\mathrm{cross} \sim 27\ \mathrm{days}$, implying $t_\mathrm{crush} \sim 2.4\ \mathrm{yr}$ and $t_\mathrm{airmass} \sim 75\ \mathrm{yr}$. This timescale is short compared with the orbital time, and so is consistent with the idea that drag on the cloud may alter its orbit, doing so most rapidly near pericenter.

Assuming that G2's orbit will morph into G1's, or at least that the two orbits are close enough to each other for G1 to have potentially accelerated the gas in G2's path, we can also compute an approximate value of $\delta$, i.e. the separation of the two clouds in units of cloud radii. Various fits \citep{pfuhl_galactic_2015, mccourt_going_2016} estimate that G1 and G2 have pericenters about 13 years apart, so roughly $\delta \sim 13\ \mathrm{yr}\ \cdot v_\mathrm{peri}/r_c \sim 190$. This is substantially less than the value of $\chi$ estimated above, making it plausible that G2 is close enough behind G1 in its orbit (so long as the orbital planes themselves are close enough) to be affected by G1. It follows that $t_\mathrm{airmass}$ may be superseded in importance by $t_\mathrm{visc}$.

If we assume the viscosity given by Equation \eqref{eq:spitzer}, i.e. $\mu \propto T^{5/2}$, and extrapolate the temperature profile assumed in \citet{steinberg_probing_2018} down to pericenter, we arrive at an incredibly short estimate of $t_\mathrm{visc}$. In particular, $r_c^2/\nu \sim 3.2 \mathrm{hr}$, far shorter than any other timescale in the problem. If the viscous time were truly this short, there would be no way for the inter-cloud stream to remain coherent between G1's and G2's pericenter passages, and hence no way for G1 to affect G2. However, the true viscous timescale is likely much longer for a few reasons. First, the $T\propto r^{-1}$ profile yields a, perhaps implausibly, high temperature of $1.25 \times 10^9\ \mathrm{K}$ for the background gas at pericenter. If the background temperature were only $10^8 K$, $r_c^2/\nu$ would be nearly 3 orders of magnitude greater ($\sim 73 \mathrm{days}$) owing to the steep temperature dependence of $\mu$. Second, as we saw in Figure \ref{fig:tvisc}, $r_c^2/\nu$ can substantially underestimate the true viscous timescale. Assuming $\chi \sim 10^3$ and $\delta \sim 200$, the effective overdensity of the stream is $\chi_\mathrm{eff} \sim 8$. Reading off the correction factor from Figure \ref{fig:tvisc} for the constant $\mu$ case, $t_\mathrm{visc}$ should be about 30 times $r_c^2/\nu$. Finally, the factor of $f_\mu<1$ in Equation \eqref{eq:spitzer} reminds us that the geometry of the magnetic field may conspire to reduce $\mu$ below the value we have assumed.

Many things have already been included in the joint modeling of the dynamics of G1 and G2, from the background gas density and its profile to the large-scale rotation or outflow of the background gas, none of which is extremely well-constrained. Flipping the problem to employ the observed trajectories of these clouds to constrain the gas properties appears to be a promising approach \citep{mccourt_going_2016, madigan_using_2017}. Here we suggest that ultimately the mutual hydrodynamic interaction of these clouds and any associated stream of gas may play an important role in these dynamics, and may constrain the temperature profile of gas near Sgr A$^*$ via the viscosity of the background gas.

\subsubsection{The Circumgalactic Medium}
\begin{figure*}
\includegraphics[width=17cm]{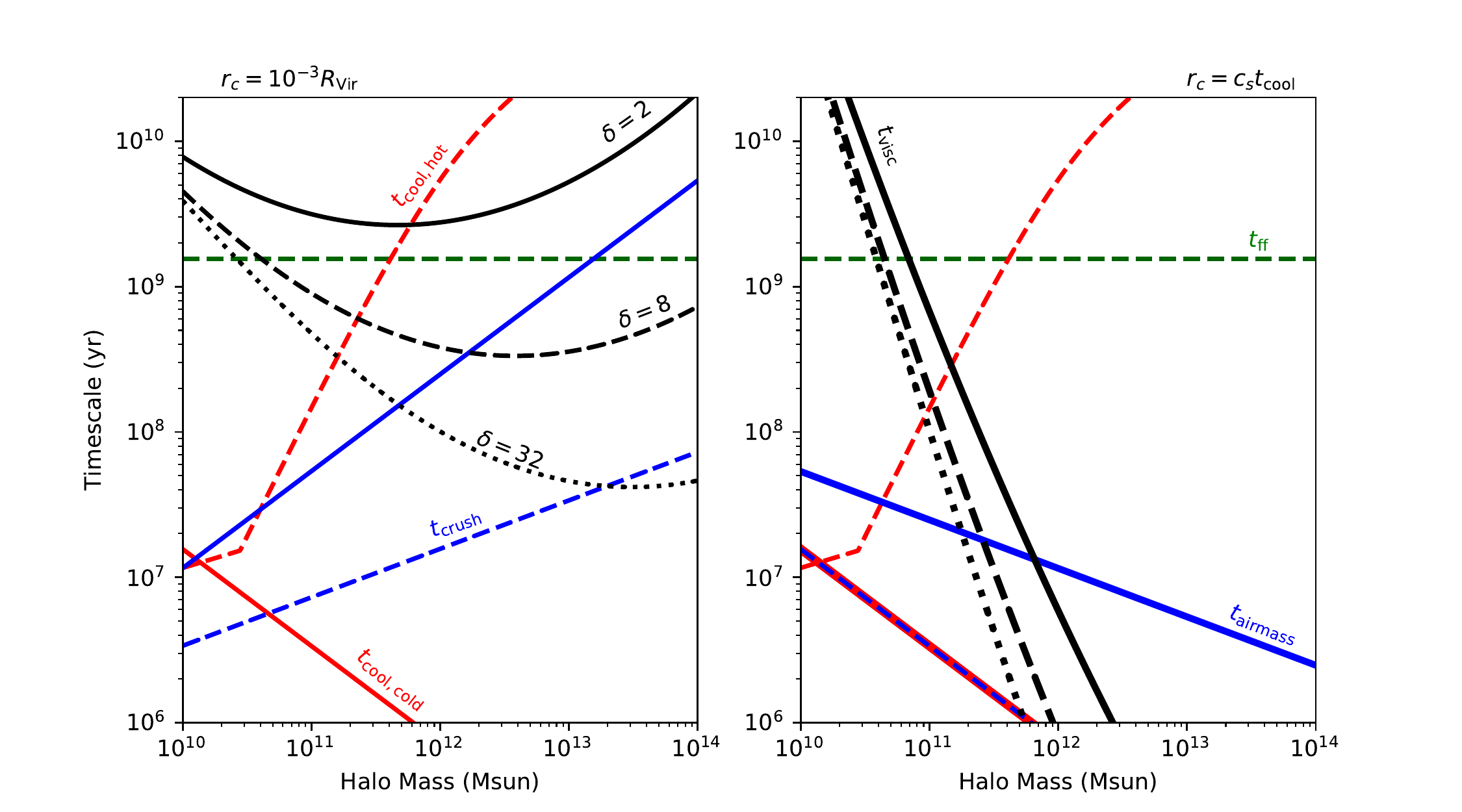}
\caption{Estimates of various timescales governing the dynamics of cold clouds in shock-heated halo gas. The left panel shows these timescales when the cloud sizes are proportional to the Virial radius, and the right panel shows the same timescales when the cloud size is equal to the characteristic cooling scale proposed by \citet{mccourt_characteristic_2018}. }
\label{fig:halo}
\end{figure*}

Multiphase gas is routinely observed in the diffuse medium surrounding galaxies, extending out to scales comparable to the Virial radius \citep{tumlinson_cos-halos_2013, werk_cos-halos_2014, bordoloi_cos-dwarfs_2014}. In the Milky Way and nearby galaxies, we can see a component of this gas in HI as High Velocity Clouds \citep[e.g.][]{putman_hipass_2002, heitsch_fate_2009} and their extragalactic analogues \citep{sancisi_cold_2008}, as well as the Magellanic Stream, an interwoven tail of gas originating from the Magellanic Clouds \citep[see][for a review]{donghia_magellanic_2016}.

Aside from clear cases like the Magellanic Stream, the origin of this cold gas is not always clear, especially for cases where it is only seen in absorption along a single sightline. A substantial portion of cosmological accretion is expected to occur in cold streams \citep[e.g.][]{keres_how_2005, dekel_cold_2009, nelson_impact_2015}. Outflows may also play a substantial role in producing cold gas at scales comparable to the Virial radius \citep{hummels_constraints_2013, liang_column_2016, fielding_impact_2017}, though \citet{thompson_origin_2016} has suggested that entraining cold gas in the ISM and transporting it to large radii is generically difficult because of the short crushing timescale for individual clouds. Gas may also cool directly out of the hot gas owing to thermal instability in some circumstances \citep[e.g.][]{balbus_general_1995,mccourt_thermal_2012}. 

The frictional dynamics of this multiphase gas is interesting for a number of reasons. First, the line-of-sight kinematics of cold gas observed in absorption is directly observable. Second, the ultimate fate of gas ejected from a galaxy may be determined by its interaction with the hot background. Third, the angular momentum of accreting gas is likely to be affected by its interaction with the halo, and finally the smoothness of the accretion on to the galaxy, which in turn affects the scatter in galaxy scaling relations \citep{forbes_origin_2014}, may well be determined by the path the accreting gas takes through the halo.

To begin to estimate where the dynamics of streams vs. individual clouds are important, we estimate the following timescales as a function of halo mass:
\begin{enumerate}
\item $t_\mathrm{cool} \sim k_B T / (n\Lambda(T))$
\item $t_\mathrm{crush} \sim (r_c/c_s) \sqrt{\chi}$
\item $t_\mathrm{airmass} \sim (r_c/c_s) \chi$
\item $t_\mathrm{dyn} \sim 1/\sqrt{G\rho}$
\item $t_\mathrm{visc} \sim (r_c^2\rho/\mu) f(\chi_\mathrm{eff}) $
\end{enumerate}
First we assume that $T$ is equal to the Virial temperature of the halo in question for the hot phase, and $10^4$ K for the cold phase. Assuming pressure equilibrium between the hot and cold phases implies that $\chi \approx T_\mathrm{Vir}/(10^4\ \mathrm{K})$. We assume that the clouds move with a typical velocity equal to the sound speed in the hot medium, which is also of order the circular velocity of the halo. We also need to estimate the number density $n$. To do so we make the extremely crude approximation that some fraction of order unity of a given halo's universal baryon content, i.e. $f_b M_h$, with the univeral baryon fraction $f_b \approx 0.16$ \citep{planck_collaboration_planck_2016}, will be present in the hot phase. This ignores the possibility that feedback may eject gas beyond the Virial radius, the likely density variation necessary to keep such a halo in hydrostatic equilibrium, and any consideration of how quickly this putative hot gas would cool, which can be accounted for in more detailed modeling \citep[e.g.][]{lin_two-phase_2000, maller_multiphase_2004, faerman_massive_2017}. The viscous time depends on the effective density contrast as per Figure \ref{fig:tvisc}, which we include using the fitting formula, equation \eqref{eq:tviscFit}

Finally, the dynamics of clouds in this environment depends critically on the assumed cloud sizes $r_c$. We plot each of the timescales in Figure \ref{fig:halo} as a function of halo mass, and with two different assumed cloud sizes. In the left panel, we assume that the clouds are a fixed fraction of the Virial radius. This is likely a reasonable proxy for accretion flows and streams associated with dwarf galaxies, e.g. the Magellanic Stream, where the physical size of the stream is related to its cosmological origin. Another possibility, following \citet{mccourt_characteristic_2018} is that clouds, especially if they originate from thermal instability in the hot medium, may rapidly cool and fragment to a size of order $t_\mathrm{cool} c_s$. These two cases are shown in the left and right panels of Figure \ref{fig:halo} respectively.

If nothing else, this diagram demonstrates the richness of the physical processes governing the multiphase medium in galactic halos of various sizes. First, we note that regardless of the assumed cloud sizes, the cooling time in the hot medium rises dramatically as a function of halo mass - this is entirely driven by the increase in Virial temperature, since the gas density of the hot medium is, by construction, constant as a function of halo mass. Above halo masses of $\sim 10^{12} M_\odot$, the gas cools slowly enough that a stable galactic atmosphere at the Virial temperature is plausible \citep[e.g.][]{rees_cooling_1977, dekel_galaxy_2006}. The cooling time of the cold gas, meanwhile, is always quite short, and gets shorter with increasing halo mass; in order to maintain pressure balance between the phases, the cold phase gets denser with increasing halo mass, driving down the cooling time. The freefall time is also roughly constant, both with halo mass and of course with cloud size. By definition, the mean density within a dark matter halo at the Virial radius is a fixed multiple of the critical density of the universe, typically of order a tenth of a Hubble time.

Just like the cooling time, the cloud crushing time tends to be short compared to the freefall time (in fact the crushing time is equal to the cooling time when $r_c=c_s t_\mathrm{cool}$ and the clouds move at $\mathcal{M}=1$ with respect to the hot medium). The shortness of the crushing time leads to the expectation that cold clouds should be hydrodynamically disrupted, as indeed happens in our simulations. However, given that the cooling time is far shorter than $t_\mathrm{crush}$ when $r_c = 10^{-3} R_\mathrm{Vir}$, it seems plausible that much of the gas should remain cold, even if the individual clouds are disrupted. 

Assuming that individual cold clouds persist despite the short crushing time owing to the short cooling time, their dynamics are then governed by the freefall time, and either the airmass time or the viscous time, depending on whether the clouds are isolated or behaving as a collective stream. Figure \ref{fig:halo} suggests that, in the regime where hot halos are plausible, the viscous time and freefall time are comparable, with high filling-factor streams being minimally affected by drag, even if the airmass times of individual clouds are less than the freefall time. Interestingly, lower-filling-factor streams, i.e. those with $\delta \gg 1$, may have viscous times shorter than their airmass times. This suggests such streams would slow down and be dissipated by lateral motions, at which point they may then behave as individual clouds.

\begin{figure}
\includegraphics[width=9cm]{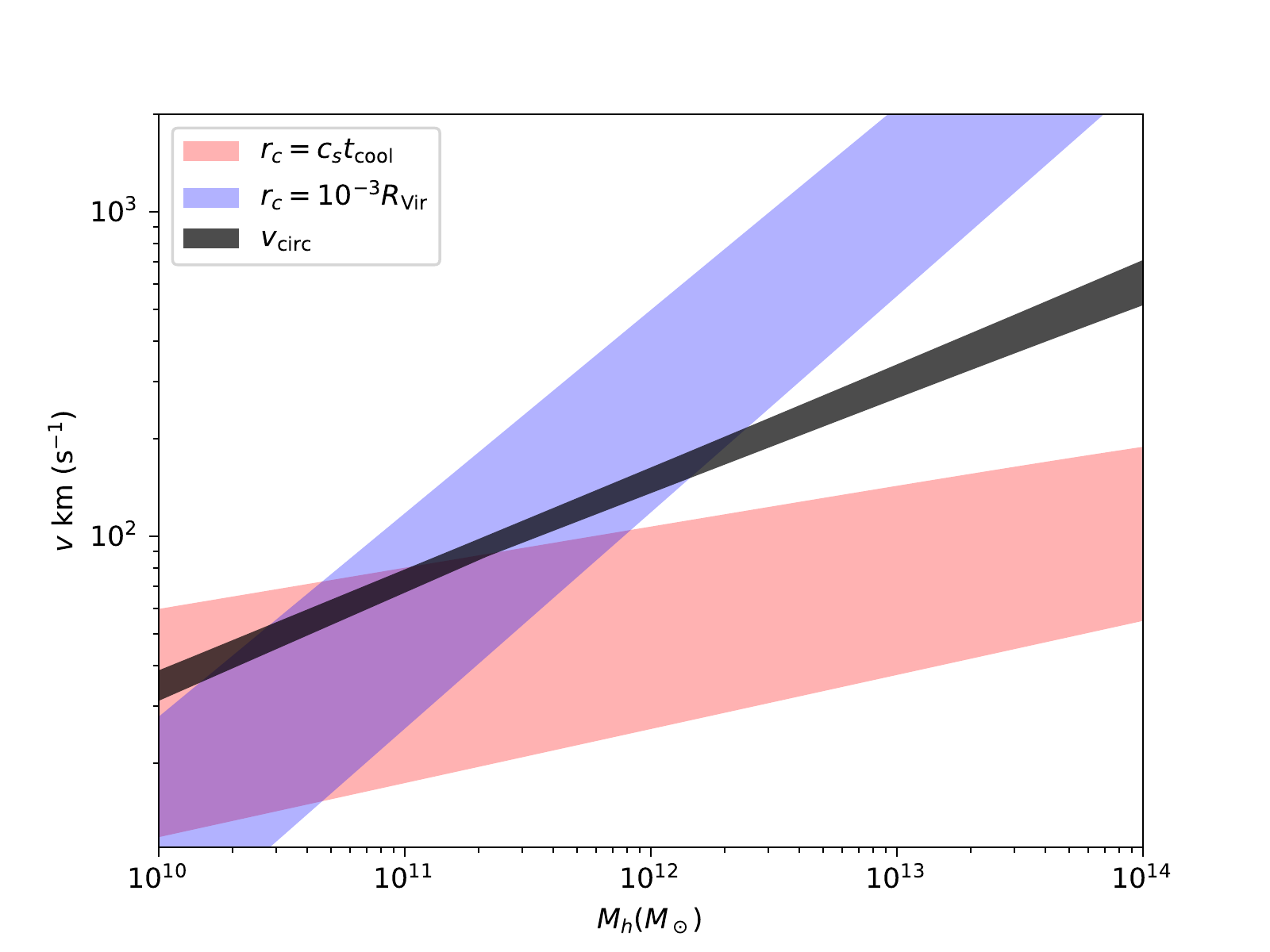}
\caption{Estimates of the terminal velocity. The range of terminal velocities for two different cloud sizes (cloud size proportional to the Virial radius in blue, and cloud size equal to the \citet{mccourt_characteristic_2018} characteristic cooling scale in red) is shown as a function of halo mass. For comparison, the circular velocity is shown in black.}
\label{fig:vTerminal}
\end{figure}

If individual clouds fragment to the sizes posited by \citet{mccourt_characteristic_2018}, the airmass time and viscous times both become incredibly short, suggesting that they should be very well-coupled with the background gas in their immediate vicinity. If the clouds are acting individually, they will quickly approach their terminal velocity. Making the same assumptions as used to construct Figure \ref{fig:halo}, we can estimate the terminal velocities of these clouds by setting the drag force $\sim (1/2) \pi r_c^2 \rho_h v^2$ equal to the gravitational force $GM(<r)(4/3)\pi r_c^3/r^2$, where $M(<r)$ is the halo mass contained within a halo-centric radius $r$ and solving for $r$. The dependence on $r$ implies there will be a range of terminal velocities at a fixed halo mass, as shown in Figure \ref{fig:vTerminal} for both of our assumptions about $r_c$. 

The terminal velocities when $r_c=10^{-3} R_\mathrm{Vir}$ are generally substantially larger than the circular velocities of the halo in the part of the diagram where we expect hot halos to exist, consistent with the fact that the airmass time in the left panel of Figure \ref{fig:halo} approaches the freefall time. In other words, in this regime, cold clouds are likely minimally affected by drag, and freefall into the galaxy. Meanwhile individual clouds at the cooling scale, i.e. for $r_c = t_\mathrm{cool} c_s$ are slowed to velocities appreciably less than the circular velocity. Measurements of the kinematics of low ionization absorbers in quasar sightlines passing near foreground galaxies would observe a range of velocities substantially smaller than the expected escape speed. This does not appear to be the case in the COS halos data \citep{tumlinson_cos-halos_2013}, suggesting that either clouds do not fragment to this scale, or if they do, they act collectively to avoid being slowed down. A more rigorous exploration of this dataset may be worthwhile.

\section{Summary}

In this work we have explored the dynamics of streams of cold clouds traveling relative to a homogenous uniform background gas close to pressure equilibrium. The standard picture of these dynamics holds that individual clouds are shredded by their interaction with the background medium on a timescale of order the cloud crushing time, $t_\mathrm{crush} = t_\mathrm{cross}\sqrt{\chi}$. If the clouds are not shredded, they are accelerated such that their velocity approaches the velocity of the background gas on a timescale of order $t_\mathrm{accel} = t_\mathrm{cross}\chi$.

When individual clouds happen to be aligned with each other into streams, as the result of cosmological accretion, a common origin or orbit, or even by the action of magnetic fields, this picture changes in a number of ways. We have explicitly shown with idealized 3D adaptive mesh refinement simulations that the clouds in these streams can shield themselves from disruption to some degree if they are sufficiently close together, and that this shielding also slows down the rate at which the clouds' velocities approach the background velocity.

To model the dynamics, we have shown that the velocities of the clouds after 2 acceleration times is crudely approximated by assuming that the initial momentum of the clouds is shared with a cylinder $\sim 5$ times the initial radii of the clouds. Looking in more detail at the velocity structure of the simulations, it becomes clear that the fall-off in velocity is not sudden, as one would infer from the cylinder model, but much more gradual. This behavior, both in space and time, is well-approximated by a simple 1D advection-diffusion equation. This equation can be applied to predict the timescale over which viscous diffusion of momentum acts to accelerate the cold material depending on the effective density contrast of the stream of clouds,  whether the clouds mix with the background material or not, and the functional form of the viscosity.

We examine the meaning of these results for a few specific astrophysical cases. First, we consider the G1 and G2 clouds orbiting in the central few milliparsecs of the Milky Way. If the orbits' of the clouds are sufficiently close to each other, the clouds may act as a single stream. The temperature close to the black hole is so high that the viscous evolution timescale of this putative stream may be extremely short. We suggest that strong constraints may be placed on the temperature of the hot background gas by accounting for this effect in the joint modelling of the orbits of G1 and G2.

We also construct a toy model for the circumgalactic medium, and consider the dynamics of cold clouds in an atmosphere heated to the Virial temperature of halos of various masses. We find that streams with high volume filling factors should pass through the halo minimally affected by drag, even in situations where individual clouds would be slowed. Meanwhile, if clouds break up into droplets of order $r_c = t_\mathrm{cool} c_s$ as suggested by \citet{mccourt_characteristic_2018}, they must behave collectively, i.e. as a larger cloud, to avoid being slowed to very modest terminal velocities likely inconsistent with absorption line observations.

\section*{Acknowledgements}

JCF is supported by an ITC Fellowship. Simulations were carried out on the UCSC supercomputer Hyades supported by NSF grant AST-1229745, and in part on the Odyssey cluster supported by the FAS Division of Science, Research Computing Group at Harvard University. The authors would like to thank Mark Krumholz, J.X. Prochaska, Evan Schneider, Cameron Liang, Stephen Murray, Mike McCourt, James Guillochon, and Nir Mandelker for thoughtful discussions. 

\bibliography{/Users/jforbes/updatingzotlib}

\begin{thebibliography}{}

\bibitem[\protect\citeauthoryear{Anninos, Fragile, Wilson \& Murray}{Anninos
  et~al.}{2012}]{anninos_three-dimensional_2012}
Anninos P.,  Fragile P.~C.,  Wilson J.,    Murray S.~D.,  2012, The
  Astrophysical Journal, 759, 132

\bibitem[\protect\citeauthoryear{Armillotta, Fraternali \&
  Marinacci}{Armillotta et~al.}{2016}]{armillotta_efficiency_2016}
Armillotta L.,  Fraternali F.,    Marinacci F.,  2016, Monthly Notices of the
  Royal Astronomical Society, 462, 4157

\bibitem[\protect\citeauthoryear{Armillotta, Fraternali, Werk, Prochaska \&
  Marinacci}{Armillotta et~al.}{2017}]{armillotta_survival_2017}
Armillotta L.,  Fraternali F.,  Werk J.~K.,  Prochaska J.~X.,    Marinacci F.,
  2017, Monthly Notices of the Royal Astronomical Society, 470, 114

\bibitem[\protect\citeauthoryear{Balbus}{Balbus}{1995}]{balbus_general_1995}
Balbus S.~A.,  1995, The Astrophysical Journal, 453, 380

\bibitem[\protect\citeauthoryear{{Banda-Barrag\'an}, Federrath, Crocker \&
  Bicknell}{{Banda-Barrag\'an} et~al.}{2018}]{banda-barragan_filament_2018}
{Banda-Barrag\'an} W.~E.,  Federrath C.,  Crocker R.~M.,    Bicknell G.~V.,
  2018, Monthly Notices of the Royal Astronomical Society, 473, 3454

\bibitem[\protect\citeauthoryear{{Banda-Barrag\'an}, Parkin, Federrath, Crocker
  \& Bicknell}{{Banda-Barrag\'an} et~al.}{2016}]{banda-barragan_filament_2016}
{Banda-Barrag\'an} W.~E.,  Parkin E.~R.,  Federrath C.,  Crocker R.~M.,
  Bicknell G.~V.,  2016, Monthly Notices of the Royal Astronomical Society,
  455, 1309

\bibitem[\protect\citeauthoryear{Bordoloi et~al.,}{Bordoloi
  et~al.}{2014}]{bordoloi_cos-dwarfs_2014}
Bordoloi R.  et~al., 2014, The Astrophysical Journal, 796, 136

\bibitem[\protect\citeauthoryear{Bower et~al.,}{Bower
  et~al.}{2015}]{bower_radio_2015}
Bower G.~C.  et~al., 2015, The Astrophysical Journal, 802, 69

\bibitem[\protect\citeauthoryear{Bryan et~al.,}{Bryan
  et~al.}{2014}]{bryan_enzo_2014}
Bryan G.~L.  et~al., 2014, The Astrophysical Journal Supplement Series, 211, 19

\bibitem[\protect\citeauthoryear{Burkert, Schartmann, Alig, Gillessen, Genzel,
  Fritz \& Eisenhauer}{Burkert et~al.}{2012}]{burkert_physics_2012}
Burkert A.,  Schartmann M.,  Alig C.,  Gillessen S.,  Genzel R.,  Fritz T.~K.,
    Eisenhauer F.,  2012, The Astrophysical Journal, 750, 58

\bibitem[\protect\citeauthoryear{Cl\'enet, Rouan, Gratadour, Marco, L\'ena,
  Ageorges \& Gendron}{Cl\'enet et~al.}{2005}]{clenet_dual_2005}
Cl\'enet Y.,  Rouan D.,  Gratadour D.,  Marco O.,  L\'ena P.,  Ageorges N.,
  Gendron E.,  2005, Astronomy and Astrophysics, 439, L9

\bibitem[\protect\citeauthoryear{Dekel \& Birnboim}{Dekel \&
  Birnboim}{2006}]{dekel_galaxy_2006}
Dekel A.,  Birnboim Y.,  2006, Monthly Notices of the Royal Astronomical
  Society, 368, 2

\bibitem[\protect\citeauthoryear{Dekel et~al.,}{Dekel
  et~al.}{2009}]{dekel_cold_2009}
Dekel A.  et~al., 2009, Nature, 457, 451

\bibitem[\protect\citeauthoryear{D'Onghia \& Fox}{D'Onghia \&
  Fox}{2016}]{donghia_magellanic_2016}
D'Onghia E.,  Fox A.~J.,  2016, Annual Review of Astronomy and Astrophysics,
  54, 363

\bibitem[\protect\citeauthoryear{Faerman, Sternberg \& McKee}{Faerman
  et~al.}{2017}]{faerman_massive_2017}
Faerman Y.,  Sternberg A.,    McKee C.~F.,  2017, The Astrophysical Journal,
  835, 52

\bibitem[\protect\citeauthoryear{Field}{Field}{1965}]{field_thermal_1965}
Field G.~B.,  1965, The Astrophysical Journal, 142, 531

\bibitem[\protect\citeauthoryear{Fielding, Quataert, McCourt \&
  Thompson}{Fielding et~al.}{2017}]{fielding_impact_2017}
Fielding D.,  Quataert E.,  McCourt M.,    Thompson T.~A.,  2017, Monthly
  Notices of the Royal Astronomical Society, 466, 3810

\bibitem[\protect\citeauthoryear{Forbes, Krumholz, Burkert \& Dekel}{Forbes
  et~al.}{2014}]{forbes_origin_2014}
Forbes J.~C.,  Krumholz M.~R.,  Burkert A.,    Dekel A.,  2014, Monthly Notices
  of the Royal Astronomical Society, 443, 168

\bibitem[\protect\citeauthoryear{Fragile, Anninos, Gustafson \& Murray}{Fragile
  et~al.}{2005}]{fragile_magnetohydrodynamic_2005}
Fragile P.~C.,  Anninos P.,  Gustafson K.,    Murray S.~D.,  2005, The
  Astrophysical Journal, 619, 327

\bibitem[\protect\citeauthoryear{Gillessen et~al.,}{Gillessen
  et~al.}{2012}]{gillessen_gas_2012}
Gillessen S.  et~al., 2012, Nature, 481, 51

\bibitem[\protect\citeauthoryear{Heitsch \& Putman}{Heitsch \&
  Putman}{2009}]{heitsch_fate_2009}
Heitsch F.,  Putman M.~E.,  2009, The Astrophysical Journal, 698, 1485

\bibitem[\protect\citeauthoryear{Huang, Zhou \& Lin}{Huang
  et~al.}{2013}]{huang_coagulation_2013}
Huang X.,  Zhou T.,    Lin D. N.~C.,  2013, The Astrophysical Journal, 769, 23

\bibitem[\protect\citeauthoryear{Hummels, Bryan, Smith \& Turk}{Hummels
  et~al.}{2013}]{hummels_constraints_2013}
Hummels C.~B.,  Bryan G.~L.,  Smith B.~D.,    Turk M.~J.,  2013, Monthly
  Notices of the Royal Astronomical Society, 430, 1548

\bibitem[\protect\citeauthoryear{Kere{\v s}, Katz, Weinberg \& Dav\'e}{Kere{\v
  s} et~al.}{2005}]{keres_how_2005}
Kere{\v s} D.,  Katz N.,  Weinberg D.~H.,    Dav\'e R.,  2005, Monthly Notices
  of the Royal Astronomical Society, 363, 2

\bibitem[\protect\citeauthoryear{Klein, McKee \& Colella}{Klein
  et~al.}{1994}]{klein_hydrodynamic_1994}
Klein R.~I.,  McKee C.~F.,    Colella P.,  1994, The Astrophysical Journal,
  420, 213

\bibitem[\protect\citeauthoryear{Li \& Bryan}{Li \&
  Bryan}{2014}]{li_modeling_2014}
Li Y.,  Bryan G.~L.,  2014, The Astrophysical Journal, 789, 153

\bibitem[\protect\citeauthoryear{Liang, Kravtsov \& Agertz}{Liang
  et~al.}{2016}]{liang_column_2016}
Liang C.~J.,  Kravtsov A.~V.,    Agertz O.,  2016, Monthly Notices of the Royal
  Astronomical Society, 458, 1164

\bibitem[\protect\citeauthoryear{Lin \& Murray}{Lin \&
  Murray}{2000}]{lin_two-phase_2000}
Lin D. N.~C.,  Murray S.~D.,  2000, The Astrophysical Journal, 540, 170

\bibitem[\protect\citeauthoryear{McCourt \& Madigan}{McCourt \&
  Madigan}{2016}]{mccourt_going_2016}
McCourt M.,  Madigan A.-M.,  2016, Monthly Notices of the Royal Astronomical
  Society, 455, 2187

\bibitem[\protect\citeauthoryear{McCourt, Oh, O'Leary \& Madigan}{McCourt
  et~al.}{2018}]{mccourt_characteristic_2018}
McCourt M.,  Oh S.~P.,  O'Leary R.,    Madigan A.-M.,  2018, Monthly Notices of
  the Royal Astronomical Society, 473, 5407

\bibitem[\protect\citeauthoryear{McCourt, O'Leary, Madigan \& Quataert}{McCourt
  et~al.}{2015}]{mccourt_magnetized_2015}
McCourt M.,  O'Leary R.~M.,  Madigan A.-M.,    Quataert E.,  2015, Monthly
  Notices of the Royal Astronomical Society, 449, 2

\bibitem[\protect\citeauthoryear{McCourt, Sharma, Quataert \& Parrish}{McCourt
  et~al.}{2012}]{mccourt_thermal_2012}
McCourt M.,  Sharma P.,  Quataert E.,    Parrish I.~J.,  2012, Monthly Notices
  of the Royal Astronomical Society, 419, 3319

\bibitem[\protect\citeauthoryear{Madigan, McCourt \& O'Leary}{Madigan
  et~al.}{2017}]{madigan_using_2017}
Madigan A.-M.,  McCourt M.,    O'Leary R.~M.,  2017, Monthly Notices of the
  Royal Astronomical Society, 465, 2310

\bibitem[\protect\citeauthoryear{Maller \& Bullock}{Maller \&
  Bullock}{2004}]{maller_multiphase_2004}
Maller A.~H.,  Bullock J.~S.,  2004, Monthly Notices of the Royal Astronomical
  Society, 355, 694

\bibitem[\protect\citeauthoryear{Murray, White, Blondin \& Lin}{Murray
  et~al.}{1993}]{murray_dynamical_1993}
Murray S.~D.,  White S. D.~M.,  Blondin J.~M.,    Lin D. N.~C.,  1993, The
  Astrophysical Journal, 407, 588

\bibitem[\protect\citeauthoryear{{Murray-Clay} \& Loeb}{{Murray-Clay} \&
  Loeb}{2012}]{murray-clay_disruption_2012}
{Murray-Clay} R.~A.,  Loeb A.,  2012, Nature Communications, 3, 1049

\bibitem[\protect\citeauthoryear{Nelson, Genel, Vogelsberger, Springel,
  Sijacki, Torrey \& Hernquist}{Nelson et~al.}{2015}]{nelson_impact_2015}
Nelson D.,  Genel S.,  Vogelsberger M.,  Springel V.,  Sijacki D.,  Torrey P.,
    Hernquist L.,  2015, Monthly Notices of the Royal Astronomical Society,
  448, 59

\bibitem[\protect\citeauthoryear{Pfuhl et~al.,}{Pfuhl
  et~al.}{2015}]{pfuhl_galactic_2015}
Pfuhl O.  et~al., 2015, The Astrophysical Journal, 798, 111

\bibitem[\protect\citeauthoryear{Phifer et~al.,}{Phifer
  et~al.}{2013}]{phifer_keck_2013}
Phifer K.  et~al., 2013, The Astrophysical Journal Letters, 773, L13

\bibitem[\protect\citeauthoryear{Pittard}{Pittard}{2006}]{pittard_mass-loaded_2006}
Pittard J.~M.,  2006, arXiv:astro-ph/0607310

\bibitem[\protect\citeauthoryear{{Planck Collaboration} et~al.,}{{Planck
  Collaboration} et~al.}{2016}]{planck_collaboration_planck_2016}
{Planck Collaboration} et~al., 2016, Astronomy and Astrophysics, 594, A13

\bibitem[\protect\citeauthoryear{Putman et~al.,}{Putman
  et~al.}{2002}]{putman_hipass_2002}
Putman M.~E.  et~al., 2002, The Astronomical Journal, 123, 873

\bibitem[\protect\citeauthoryear{Rees \& Ostriker}{Rees \&
  Ostriker}{1977}]{rees_cooling_1977}
Rees M.~J.,  Ostriker J.~P.,  1977, Monthly Notices of the Royal Astronomical
  Society, 179, 541

\bibitem[\protect\citeauthoryear{Roediger, Kraft, Forman, Nulsen \&
  Churazov}{Roediger et~al.}{2013}]{roediger_kelvin-helmholtz_2013}
Roediger E.,  Kraft R.~P.,  Forman W.~R.,  Nulsen P. E.~J.,    Churazov E.,
  2013, The Astrophysical Journal, 764, 60

\bibitem[\protect\citeauthoryear{Sancisi, Fraternali, Oosterloo \& {van der
  Hulst}}{Sancisi et~al.}{2008}]{sancisi_cold_2008}
Sancisi R.,  Fraternali F.,  Oosterloo T.,    {van der Hulst} T.,  2008,
  Astronomy and Astrophysics Review, 15, 189

\bibitem[\protect\citeauthoryear{Sarazin}{Sarazin}{1988}]{sarazin_x-ray_1988}
Sarazin C.~L.,  1988, X-Ray Emission from Clusters of Galaxies

\bibitem[\protect\citeauthoryear{Schneider \& Robertson}{Schneider \&
  Robertson}{2015}]{schneider_cholla_2015}
Schneider E.~E.,  Robertson B.~E.,  2015, The Astrophysical Journal Supplement
  Series, 217, 24

\bibitem[\protect\citeauthoryear{Schneider \& Robertson}{Schneider \&
  Robertson}{2017}]{schneider_hydrodynamical_2017}
Schneider E.~E.,  Robertson B.~E.,  2017, The Astrophysical Journal, 834, 144

\bibitem[\protect\citeauthoryear{Shakura \& Sunyaev}{Shakura \&
  Sunyaev}{1973}]{shakura_black_1973}
Shakura N.~I.,  Sunyaev R.~A.,  1973, Astronomy and Astrophysics, 24, 337

\bibitem[\protect\citeauthoryear{Spitzer}{Spitzer}{1962}]{spitzer_physics_1962}
Spitzer L.,  1962, Physics of {{Fully Ionized Gases}}

\bibitem[\protect\citeauthoryear{Steinberg et~al.,}{Steinberg
  et~al.}{2018}]{steinberg_probing_2018}
Steinberg E.  et~al., 2018, Monthly Notices of the Royal Astronomical Society,
  473, 1841

\bibitem[\protect\citeauthoryear{Thompson, Quataert, Zhang \&
  Weinberg}{Thompson et~al.}{2016}]{thompson_origin_2016}
Thompson T.~A.,  Quataert E.,  Zhang D.,    Weinberg D.~H.,  2016, Monthly
  Notices of the Royal Astronomical Society, 455, 1830

\bibitem[\protect\citeauthoryear{Tumlinson et~al.,}{Tumlinson
  et~al.}{2013}]{tumlinson_cos-halos_2013}
Tumlinson J.  et~al., 2013, The Astrophysical Journal, 777, 59

\bibitem[\protect\citeauthoryear{Tumlinson et~al.,}{Tumlinson
  et~al.}{2011}]{tumlinson_large_2011}
Tumlinson J.  et~al., 2011, Science, 334, 948

\bibitem[\protect\citeauthoryear{Wang}{Wang}{1995}]{wang_cooling_1995}
Wang B.,  1995, The Astrophysical Journal, 444, 590

\bibitem[\protect\citeauthoryear{Werk et~al.,}{Werk
  et~al.}{2014}]{werk_cos-halos_2014}
Werk J.~K.  et~al., 2014, arXiv:1403.0947 [astro-ph]

\bibitem[\protect\citeauthoryear{Witzel et~al.,}{Witzel
  et~al.}{2017}]{witzel_post-periapsis_2017}
Witzel G.  et~al., 2017, The Astrophysical Journal, 847, 80

\bibitem[\protect\citeauthoryear{Yuan, Quataert \& Narayan}{Yuan
  et~al.}{2003}]{yuan_nonthermal_2003}
Yuan F.,  Quataert E.,    Narayan R.,  2003, The Astrophysical Journal, 598,
  301

\end{thebibliography}

\clearpage

\end{document}